\documentstyle[aps,prl,epsf,floats]{revtex}
\begin{document}
\twocolumn[\hsize\textwidth\columnwidth\hsize\csname@twocolumnfalse%
\endcsname

\title{Charge and orbital order in half-doped manganites}
\author{Jeroen van den Brink, Giniyat Khaliullin~\cite{Kazan}}
\address{
Max-Planck-Institut f\"ur Festk\"orperforschung,
Heisenbergstr. 1, D-70569 Stuttgart, Germany }
\author{Daniel Khomskii}
\address{
Laboratory of Applied and Solid State Physics, Materials Science Center,
University of Groningen, Nijenborgh 4, 9747 AG Groningen, The Netherlands}
\date{\today}
\maketitle
\begin{abstract}
An explanation is given for the charge order, orbital
order and insulating state observed in half-doped manganese oxides,
such as Nd$_{1/2}$Sr$_{1/2}$MnO$_{3}$.
The competition between the kinetic energy of the
electrons and the magnetic exchange energy drives the
formation of effectively one-dimensional ferromagnetic zig-zag chains. 
Due to a topological phase factor 
in the hopping, the chains are intrinsically insulating and orbital-ordered.
Most surprisingly, the strong Coulomb interaction between electrons on the same Mn-ion
leads to the experimentally observed charge ordering.
For doping less than $1/2$ the system is unstable towards 
phase separation into a ferromagnetic metallic and charge-ordered 
insulating phase.
\end{abstract}
\pacs{PACS number(s): 75.30.Ds, 75.30.Et}]

\begin{figure}
      \epsfysize=51mm
      \centerline{\epsffile{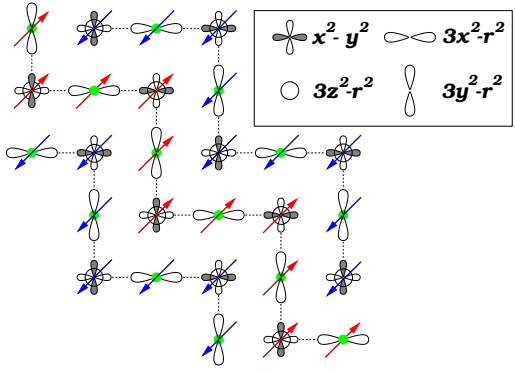}}
\caption{View of the CE-phase in the x-y plane. We choose our basis orbitals such that 
the gray lobes of the shown orbitals have a negative sign.
The dots at the bridge-sites represent a charge-surplus.}
\label{fig:ce_phase}
\end{figure}

Manganese oxides with the general composition R$_{1-x}$A$_x$MnO${_3}$ (where R and A are rare- and
alkaline-earth ions, respectively) have attracted considerable attention because of their
unusual magnetic and electronic properties. 
In some of these materials, metal-insulator
transitions can be observed where both conductivity and magnetization change markedly. 
The $x = 0$ and $x = 1$ end members of the R$_{1-x}$A$_x$MnO${_3}$ family are insulating and 
antiferromagnetic (AF)
with the Mn-ion in the Mn$^{3+}$ and Mn$^{4+}$ state, respectively. For intermediate $x$, 
the average Mn-valence is non-integer and the material is generally metallic or semiconducting. 
Most of the perovskite manganites show a ferromagnetic (FM) groundstate when the
holes are optimally doped (usually $0.2<x<0.5$) and anisotropic antiferromagnetic (AFM) phases
for $x>0.5$. The half-doped manganites, with $x=\frac{1}{2}$, are very particular. Magnetically
these systems form FM zig-zag chains that are
coupled AFM (see Fig.\ref{fig:ce_phase}) at low temperatures, the so-called magnetic CE-phase~\cite{Wohlan55}.
The groundstate is, moreover, an orbital-ordered and charge-ordered insulator. This
behavior is generic and is experimentally observed in 
Nd$_{1/2}$Sr$_{1/2}$MnO$_3$~\cite{Kuwahara95,Kawano97},
Pr$_{1/2}$Sr$_{1/2}$MnO$_3$ ~\cite{Tomioka95},  
Pr$_{1/2}$Ca$_{1/2}$MnO$_3$~\cite{Tomioka96}, 
La$_{1/2}$Ca$_{1/2}$MnO$_3$ ~\cite{Urushibara95,Mori98},  
Nd$_{1/2}$Ca$_{1/2}$MnO$_3$~\cite{Kimura_prep}
and in the half-doped layered manganite 
La$_{1/2}$Sr$_{3/2}$MnO$_4$~\cite{Moritomo95}.
The insulating charge-ordered state can be transformed into a metallic FM state by
application of a external magnetic field, a transition that is accompanied by a change
in resistivity of several orders of magnitude~\cite{Kuwahara95,Tokunaga98}.

The occurrence of charge order, orbital order and large magneto-resistance in the half-doped 
manganites is experimentally well established. Theoretically, however, 
the nature of the charge ordering at $x=\frac{1}{2}$ and the origin of the unconventional
zig-zag magnetic structure remain unclear.
We address these points here and show that $(i)$ the insulating CE-phase results from a
particular ordering of orbitals, $(ii)$ this state is stable only in a narrow concentration
range around the commensurate value and that $(iii)$ the zig-zag chains are intrinsically
charge ordered due to on-site Hubbard correlations.

The competition between kinetic energy (double-exchange) and
superexchange between the manganese core-spins, combined with the orbital degeneracy
leads to the formation of the magnetic CE-phase, for the same reasons as it leads
to the anisotropic magnetic phases in the highly doped manganites with $x>\frac{1}{2}$~\cite{Brink99}.
The CE-phase is found to be stable close to $x=\frac{1}{2}$; for
$x>0.57$ the C-phase is stable and for $x<\frac{1}{2}$ we find phase separation between
the CE-phase and FM-phase.

As in the double-exchange framework electrons can only hop between sites with FM aligned
core-spins, in the CE-phase only hopping processes within the zig-zag chains are
possible, rendering the system one-dimensional for low-energy charge fluctuations. 
The essential observation is that an electron that passes a corner-site of the zig-zag 
chain, acquires a phase that
depends on the orbital through which it passes. This leads to an effective 
dimerization that splits the bands and opens a gap at the Fermi-surface. The gap is
very robust as it is a consequence of the staggered phase-factor that is itself
fully determined by the topology of the system. At the same time not all orbitals
are fully occupied, leading to an orbital-polarized groundstate.
Our most surprising observation is that the experimentally observed charge order is directly obtained
from the degenerate double-exchange model when the Coulomb interaction 
(the Hubbard $U$) between electrons in different orbitals, but {\it on the same site}
is included. This can be understood from the fact that in the band-picture on the corner-sites
both orthogonal orbitals are partially occupied, but on the bridge-site only one orbital
is partially filled. The on-site Coulomb interaction acts therefore differently on the
corner- and bridge-sites: charge is pushed away from the effectively correlated
corner-sites to the effectively uncorrelated bridge-sites.

The theoretical explanation for the ferromagnetic metallic state in doped manganites was
already developed in 50's and 60's~\cite{Zener51,Anderson55,DeGennes60}. 
The number of $d$ electrons per manganese site is
$4-x$. Three electrons occupy the low lying $t_{2g}$ orbitals, having parallel spins.
The other electrons occupy the $e_g$ orbitals with their spin parallel to the
$t_{2g}$ spin because of the large ferromagnetic Hund's rule exchange $J_H$.
In the double-exchange framework electrons can only hop from one Mn-site to a neighboring one, 
when the $t_{2g}$
spins of these sites are aligned ferromagnetically because otherwise,
in the classical treatment of the $t_{2g}$ core-spin the electron would 
have to overcome an energy barrier proportional to $J_H$.

On the other hand, the conventional superexchange favors AF alignment the spins.
This leads to an intricate competition between the kinetic energy and the superexchange~\cite{DeGennes60} 
that is amplified by the two-fold degeneracy of the $e_g$ levels in a cubic crystal.
As a basis for the $e_g$ wavefunctions we take the $x^2-y^2$ ($| \chi \rangle$) and $3z^2-r^2$ ($| \zeta \rangle$)
orbitals. The hopping of the electrons between neighboring Mn-sites depends strongly on the kind
of the orbitals involved and on the direction of the bond. The Hamiltonian for the kinetic
energy is 
\begin{equation}
H_t= t \Sigma_{\alpha,\beta, \langle ij \rangle_{\Gamma}} \ \ M^{\Gamma}_{\alpha,\beta} \  
c^+_{i,\alpha} c_{j,\beta}, 
\end{equation}
where $c^+_{i,\alpha}$ creates an electron on site $i$
in orbital $\alpha$ and $\langle ij \rangle_{\Gamma}$ denotes a nearest neighbor pair along the 
$\Gamma$-direction ($\Gamma=x,y,z$) and $t$ is the hopping integral.
The hopping is constrained to nearest neighbor Mn-ions that
have FM aligned $t_{2g}$ core-spins (canted phases are not important in the doping range
we consider here.)
The matrices $M^{x}_{\alpha,\beta}$ and $M^{y}_{\alpha,\beta}$
can be found by applying the cubic symmetry operations on $M^{z}_{\alpha,\beta}$, where
$M^{z}_{| \zeta \rangle ,| \zeta \rangle }=1$ and all other matrix elements are zero. This means physically that
along the $z-$direction, for instance, electrons can only hop between $| \zeta \rangle $ orbitals
and that all other hoppings (involving at least one $| \chi \rangle $ orbital) are zero by symmetry,
whereas in the $x-$ and $y-$direction all hoppings are allowed (and related by symmetry).
The strong spatial anisotropy of the hopping, combined with the competing kinetic 
and superexchange energy $J$ can, depending on the $e_g$ bandwidth and $J$,
give rise to low dimensional spin-structures that, for instance, optimize kinetic energy by forming FM chains 
and optimize  magnetic energy by a AFM coupling of these chains 
(C-phase), or AFM coupled FM planes (A-phase)~\cite{Brink99}.

\begin{figure}
      \epsfxsize=50mm
      \centerline{\epsffile{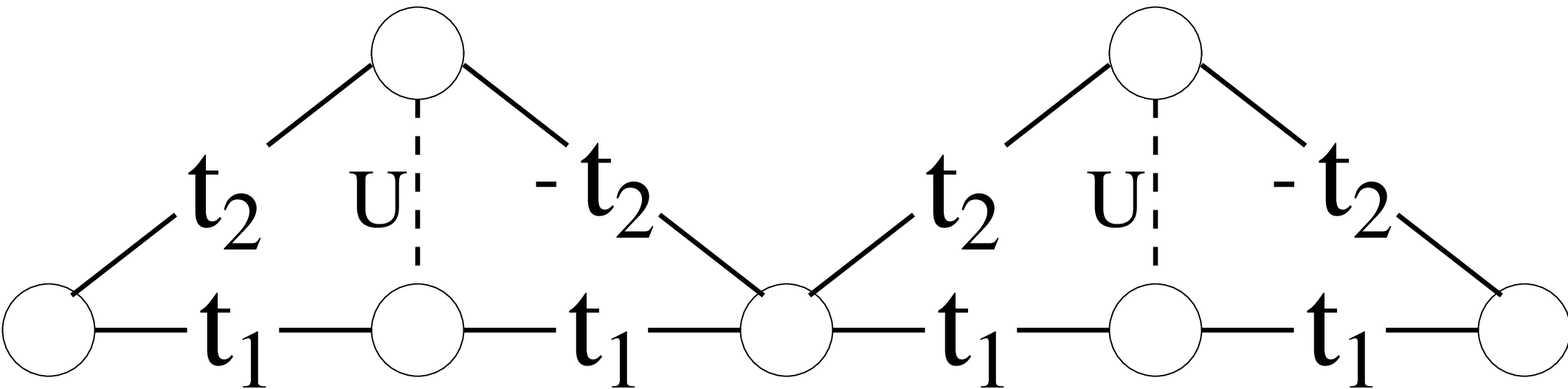}}
      \epsfxsize=85.7mm
      \centerline{\epsffile{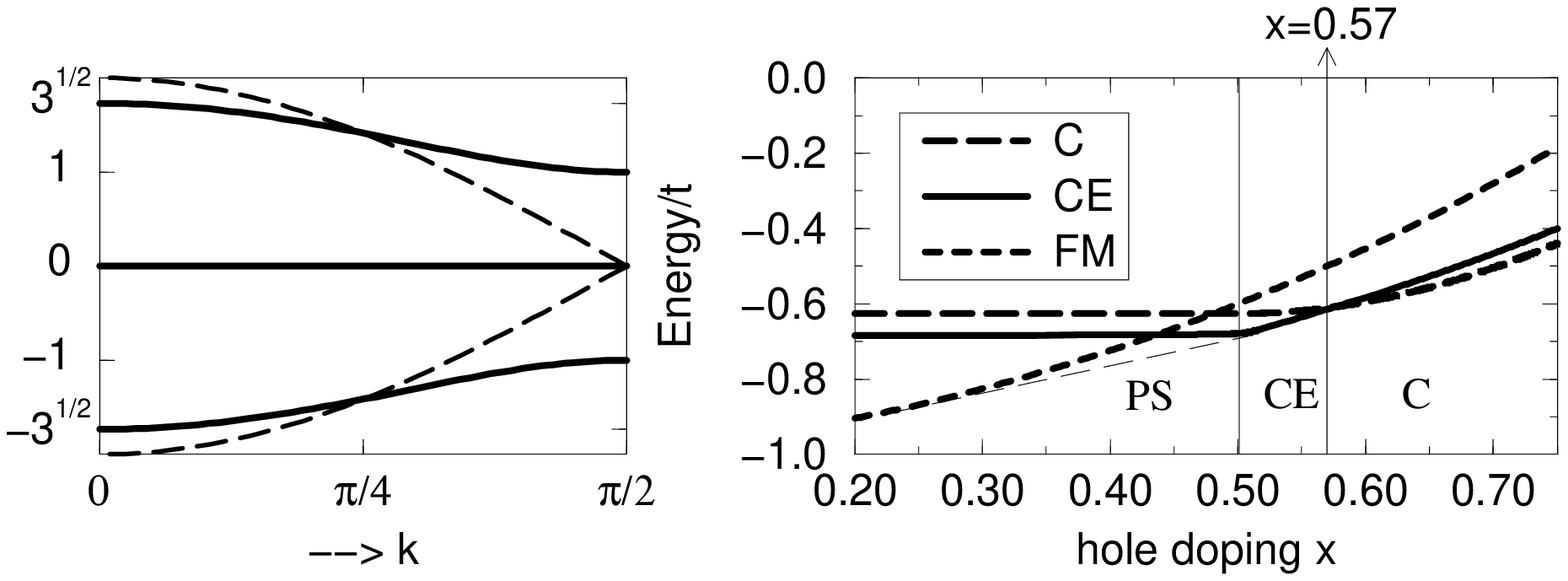}}
\caption{Top: topology of the interactions in a zig-zag chain, where $t_1=t/2$, $t_2= t \sqrt{3} /2$, 
and $U$ is the Coulomb interaction between electrons on the same site.
Bottom left: electron dispersion in the zig-zag chain of the CE-phase for $U=0$ and electron dispersion 
in a straight chain, as in C-phase (dashed line). Bottom right: total energy per site
for the CE-, C- and FM-phase for $t/J=5$. The Maxwell-construction in the phase separated (PS) region
is shown by the thin dashed line.}
\label{fig:hop}
\end{figure}

In the homogenous FM state at $x=\frac{1}{2}$ the magnetic energy per site is $E_{mag}^{FM}=3J$
and the kinetic energy per site $E_{kin}^{FM}=-t$,
if the $t_{2g}$ spin is treated as a classical spin.
Here the magnetic energy is found by determining the number of AFM bonds, and the kinetic energy
from the filling of the bands, as is described in Ref.~\cite{Brink99}. 
For AFM coupled FM chains, the C-phase, 
$E_{kin}^{C}+E_{mag}^{C}=-0.6366t+J$, so that the C-phase is stable with respect to the 
FM-phase for $J>0.1816t$. 
A third possibility is for the spins to form FM zig-zag chains that are coupled AFM, the 
CE-phase~\cite{Wohlan55}, see Fig.\ref{fig:ce_phase}. The CE-phase is always higher in energy than the C-phase, 
except close
to $x=\frac{1}{2}$. For the half-doped system $E_{kin}^{CE}+E_{mag}^{CE}=-0.695t+J$. It is easily shown
that the A-phase is higher in energy.
Thus for the half-doped manganites the experimentally observed CE-phase is indeed
lower in energy than the C-phase, and lower than the FM-phase when $J>0.1524t$. 

Let us address our main findings that the CE-phase is charge-ordered, orbital-ordered and insulating, 
and come back to the stability of the CE-phase at the end of the paper.
As shown in Fig.\ref{fig:ce_phase} the CE-phase contains two geometrically inequivalent sites,
so called bridge- and corner-sites.
Note that our specific choice of basis-orbitals as shown in Fig.\ref{fig:ce_phase}  is motivated by the
convenience of this basis for the calculations. The expectation value of actual observables
is, of course, independent from the choice of basis Wannier orbitals.
The topology of the
electron hopping integrals is  shown in Fig.\ref{fig:hop}.
The crucial observation is that an electron that hops from one bridge-site to another
bridge-site via a $| \chi \rangle $ corner-orbital obtains a phase-factor $-1$, while if the
hopping takes place via a $| \zeta \rangle $ corner-orbital, the phase-factor is $+1$. 

The bands are obtained by the solution of a 3x3 matrix.
There are two bands with energy $\epsilon_{\pm}= \pm t\sqrt{2 + \cos k}$, where
$k$ is the wave vector, and two nondispersive bands at zero energy, see Fig.\ref{fig:hop}.~\cite{Solovyev_prep}. 
At $x=\frac{1}{2}$ the $\epsilon_{-}$ band is fully occupied, and all other bands are empty.
The system is insulating as the occupied and empty bands are split by a gap $\Delta=t$.
In the C- and CE-phase the magnetic energies are equal, but the opening of
the gap at the Fermi-energy in the CE-phase lowers its energy if the bands are half filled.
This mechanism is equivalent to the situation in the lattice-Peierls problem, where the
opening of a gap stabilizes a groundstate with a lattice deformation. 

In the insulating state the average occupancy of the bridge-site $3x^2-r^2$ ($3y^2-r^2$) orbital
is $\frac{1}{2}$ while the orthogonal bridge-orbital is empty. The average occupancy of the corner-site
is also $\frac{1}{2}$, with the ratio between $| \chi \rangle $ and $| \zeta \rangle $ 
occupancy of $\sqrt{3}:1$.
The system is therefore orbital-ordered, but charge is homogeneously distributed between
corner and bridge-sites.

This changes drastically when also the Coulomb interaction $U$ between electrons on the same
site is taken into account. The Hamiltonian is
\begin{equation}
H_U= U \Sigma_{\alpha < \beta,i} \  n_{i,\alpha} n_{i,\beta}, 
\end{equation}
where 
$n_{i,\alpha}= c^+_{i,\alpha}c_{i,\alpha}$. For the $e_g$ electrons in the 
manganites $U \approx 10t$, so that the system is strongly correlated.
The correlations, however, have a very different effect on the corner and bridge-sites.
On the bridge-site, one orbital is always empty, so that the Coulomb repulsion is
ineffective, whereas on the corner-sites both orbitals are partially occupied.
The consequence is that charge is pushed away from the correlated corner-sites
to the effectively uncorrelated bridge-sites, causing charge order.

\begin{figure}
      \epsfxsize=70mm
      \centerline{\epsffile{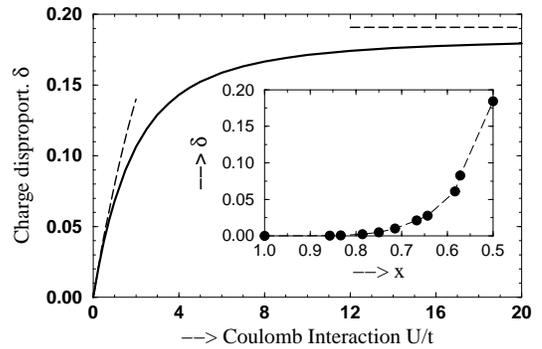}}
\caption{Charge-disproportionation as a function of Coulomb interaction and doping (inset).
The full line is obtained from exact diagonalization (ED) of a 14 site cluster. The dashed lines for 
small and large $U$ are obtained by the mean-field and Gutzwiller approximations, respectively.
The dots in the inset are the ED results for a 12 and 14 site cluster for $U \rightarrow \infty$.}
\label{fig:co}
\end{figure}

We treat the the full Hamiltonian $H=H_t+H_U$ for $x=\frac{1}{2}$ with three different
methods: exact diagonalization (ED), in mean-field (MF) and with the 
Gutzwiller-projection (GP)~\cite{Fulde91}.
In the ED calculation we consider a ring of 12 and 14 sites (or 18 and 21 orbitals, respectively) 
with the topology as in Fig.\ref{fig:hop} . In Fig.\ref{fig:co}  the charge-disproportionation $\delta$ as a function
of $U/t$ is shown, where  $\delta$ is defined as $\langle n_B-n_A \rangle$, 
where $\langle n_B \rangle$ ($\langle n_A \rangle$) is the expectation value for finding
an electron on a bridge- (corner-) site.
The results for the 12 and 14-site cluster differ by less than 2\% and finite size effects are
therefore very small.
For $U/t=0$ also $\delta=0$, as explained above, and for finite Coulomb interaction
$\delta$ monotonously increases with $U$. 
For  $U/t=10$, $\delta=0.171$ and $\delta$ reaches its maximum value for $U \rightarrow \infty$.
In a ED calculation for a 16-site cluster with  $U \rightarrow \infty$, we find
that $\delta^{ED}_{\infty}=0.185$.

For small values of $U/t$ a MF treatment is expected to be very reliable because
the charge fluctuations are gapped. 
We decouple the quartic term in Eq.2 and 
determine the charge densities in the
orbitals self-consistently. The results for small $U/t$ are shown with a dashed line
in Fig.\ref{fig:co}  and they agree very well with the ED results for  $U<\Delta$.
In the GP, valid for $U \rightarrow \infty$, we introduce constrained electrons on the correlated
corner-site. For the $| \chi \rangle$ orbital we introduce $\bar{x}_i=(1-n_i^z) x_i$ and for the
$| \zeta \rangle$ orbital $\bar{z}_i=(1-n_i^x) z_i$, where $x_i$  ($z_i$) denotes an electron
operator on corner-site $i$, with corresponding density $n_i^x$ ($n_i^z$) and 
$\bar{x}_i$ ($\bar{z}_i$) a constrained electron operator. 
The Hamiltonian is
\begin{eqnarray}
H_{GP} &=& \Sigma_{i \in B} \ c^{\dag}_i \left[ t_1 (\bar{z}_{i+1} +\bar{z}_{i-1}) \right.  \nonumber \\
&+& t_2  (\bar{x}_{i+1} - \bar{x}_{i-1}) \left. \right] +h.c.,
\end{eqnarray}
which physically means that an electron can only hop to a $| \chi \rangle$-orbital on a corner-site
when there is no electron in the $| \zeta \rangle$-orbital on that site and vice versa, so that
a corner-site can never be occupied by more than one electron, as is required in the
$U$-infinity limit. 
We decouple the quartic terms in $H_{GP}$ as $c^{\dagger}_{i+1}z_in^x_i \rightarrow 
c^{\dagger}_{i+1} z_i \langle n^x_i \rangle + \langle c^{\dagger}_{i+1} z_i \rangle n^x_i$. 
After a Fourier transform we find
\begin{eqnarray}
H_{GP}^{mf} &=& \Sigma_{k} \ \epsilon_x x^{\dag}_k x_k  
+\epsilon_z z^{\dag}_k z_k \nonumber \\
&+&  \tilde{t}_{1k} (c^{\dag}_k z_k + z^{\dag}_k c_k)+
\tilde{t}_{2k} (c^{\dag}_k x_k + x^{\dag}_k c_k)  ,
\end{eqnarray}
with
$\epsilon_x = -4 t_1  \Sigma_k  \cos{k} \ \langle  c^{\dag}_k z_k \rangle$, 
$\epsilon_z =  4 i t_2  \Sigma_k  \sin{k} \ \langle c^{\dag}_k x_k \rangle$,
$\tilde{t}_{1k} = 2 t_1  \langle 1-n^x \rangle \cos{k}$ and 
$\tilde{t}_{2k} = -2 i t_2  \langle 1-n^z \rangle \sin{k}$. 
We obtain the solution of this system of equations by iteration, and find 
the charge-disproportionation $\delta^{GP}_{\infty}=0.191$, which is represented by the dashed line 
in Fig.\ref{fig:co} for large $U/t$, and agrees well with the ED results.

The on-site Coulomb interaction causes a charge-surplus on the bridge-sites in the x-y plane. In the CE-structure the zig-zag
chains are stacked AFM along the z-direction, which implies that above each bridge-site there is 
another bridge-site in the next plane. So we find that the charges actually accumulate
on sheets formed by the bridge-sites along the z-direction. This is in remarkable agreement with 
experiment and at the same time excludes the possibility that the charge-order is driven by longer 
range Coulomb interactions because the Madelung-sum is always minimized for a 
rocksalt-type charge-order. Similar physics may also apply
to the situation at $x>0.5$, e.g. for the stripe phases observed in Ref.~\cite{Mori98}.

Fig.\ref{fig:co}  shows that the charge disproportionation is strongly doping dependent. 
For $x>\frac{1}{2}$ the holes that are doped into the
lower $\epsilon_-$-band efficiently suppress charge-order. In this doping range, however,
the CE-phase becomes unstable with respect to the C-phase. 
In Fig.\ref{fig:hop}  is shown that the 
kinetic energy of the C-phase is lower for $x>0.57$. For $x<\frac{1}{2}$ the
energy per site of the CE-phase, $E_{CE}$, is constant because the extra electrons are doped
in the non-dispersive bands at zero energy, which causes a kink of $E_{CE}$ at $x=\frac{1}{2}$.
For lower hole-doping (higher electron concentration) the homogeneous FM-phase is lower
in energy, as is expected. 
The stability of the phases can be tested by performing a Maxwell-construction.
Due to the kink in $E_{CE}$ at $x=\frac{1}{2}$, one finds that for $x<\frac{1}{2}$ the FM-phase and CE-phase
coexist. The doping region where this phase separation occurs depends strongly on the
actual value of the electron bandwidth and magnetic exchange. 
We conclude that there is a strong asymmetry for doping lower and higher than $\frac{1}{2}$,
as phase separation between the CE- and FM-phases is only present on the lower doping side.
Inclusion of the correlations beyond the mean-field approximation and longer range Coulomb 
interactions, which we did not consider in the calculation of total
energies, may modify these results. 
On the basis of the ED results we believe, however, that $U$
has only a small effect on the total energies. A more relevant contribution to the
total energy comes from lattice deformations, as it is obvious that in the Jahn-Teller distorted
C- and, 
to a lesser extend,
CE-phases~\cite{JT_CE} there is a large gain in lattice energy with respect to the undistorted
FM-phase. It can be shown that the lattice contributions to the groundstate energy
can, to a large extend, be described by using effective values of $J$~\cite{Brink_tbp},
so that our general conclusions about phase separation for $x<\frac{1}{2}$ and
about the asymmetry of the phase diagram for $x<\frac{1}{2}$ and $x>\frac{1}{2}$
remain valid. Recently this phase separation into a FM metallic and a charge ordered AFM insulating
phase was observed experimentally~\cite{Babushkina99}.

In summary, we have given an explanation for the charge order, orbital
order and insulating state of the half-doped manganese oxides.
The CE-phase with one-dimensional ferromagnetic zig-zag chains is stable for $x=\frac{1}{2}$ and 
due to staggered phase factors in the hopping, the chains are insulating and orbital-ordered.
The striking feature of our model is that the strong Coulomb interaction between electrons 
on the same Mn-site leads to the experimentally observed charge ordering.
In a magnetic field we expect the chains to be unstable
with respect to the ferromagnetic metallic state. This might 
offer a likely explanation for the large magneto-resistance at the metal-insulator transition
for the half-doped manganites.

We thank I. Solovyev and P. Horsch for useful discussions.
J.v.d.B. acknowledges with appreciation the support by the Alexander
von Humboldt-Stiftung, Germany.
This work was financially supported by the Nederlandse Stichting voor Fundamenteel
Onderzoek der Materie (FOM) and by the European network OXSEN.


\begin{thebibliography}{99}
\bibitem{JT_CE} Note that the orbital polarization on the corner-sites in the CE-phase is not complete.
\end{thebibliography}
\end{document}